\documentstyle[psfig,12pt]{article}
\newcommand{\h}{\hspace*{5 ex}}
 
\begin{document}
 

\centerline{\large\bf ABOUT SOME REGGE--LIKE RELATIONS}
\vspace*{2mm}
\centerline{\large\bf FOR (STABLE) BLACK HOLES}
 
 
\vspace*{0.5 cm}
 
\centerline{E. Recami$^{(*)}$, V. Tonin-Zanchin$^{(**)}$, A. Del Popolo$^{(***)}$ and M. Gambera$^{(***)}$}
\footnotetext{$^{(*)}${\em Facolt\`{a} di Ingegeneria, Universit\`{a}
Statale di Bergamo, Dalmine (BG), Italy.}}
\footnotetext{$^{(**)}$ {\em Dept. of Applied Mathematics, State University at
Campinas, S.P., BRAZIL.}}
\footnotetext{$^{(**)}$ {\em Osservatorio Astrofisico di Catania and CNR-GNA, 
Viale A.Doria, 6 - I 95125 Catania, ITALY (e-mail: mga@sunct.ct.astro.it)}}
\vspace*{0.1 cm}
{\small
\begin{center}
{$^{(*)}$\em and Facolt\`{a} di Ingegeneria, Universit\`{a}
Statale di Bergamo, Dalmine (BG), ITALY.}\\
\vspace*{0.1 cm}
{$^{(**)}$\em Dept. of Applied Mathematics, State University at
Campinas, S.P., BRAZIL.}\\
\vspace*{0.1 cm}
{$^{(***)}$ \em Istituto di Astronomia dell'Universit\`a di Catania, Viale A.Doria 6,\\
 I 95125 Catania, ITALY}
\end{center}
}

\vspace*{0.8 cm}
\begin{center}
{\bf ABSTRACT}
\end{center}
 
{\small
~\\ 
We associated, in a classical formulation of
``strong gravity", hadron constituents with suitable stationary, 
axisymmetric solutions of some new Einstein--type equations supposed 
to describe the strong field inside hadrons. 
These new equations can be obtained by the Einstein equations 
with cosmological term $\Lambda$.
As a consequence, $\Lambda$ and the masses $M$ result in
our theory to be scaled up, and transformed into a ``hadronic
constant" and into ``strong masses", respectively. Due to the unusual
range of $\Lambda$ and $M$ values considered, we met a series of
solutions of the Kerr--Newman--de Sitter (hereafter KNdS) type with rather
interesting properties.\\
The requirement that those solutions be stable, i.e., that
their temperature (or surface gravity) be {\em vanishingly small}, implies
the coincidence of at least two of their (in general, three) horizons.
Imposing the stability condition of a certain horizon does yield (once
chosen the values of $J, q$ and $\Lambda$) mass and radius of the
associated black--hole (hereafter BH).\\
In the case of ordinary Einstein equations and for
stable BHs of the KNdS type, we get in particular {\sc Regge--like} (hereafter 
RL) relations among mass $M$, angular momentum $J$, charge $q$ and cosmological
constant $\Lambda$; which did not receive enough attention in the previous
literature. 
Besides, we show some particular and interesting cases of these relations.\\
Another interesting point is that, with few
exceptions, all such relations (among $M, J, q, \Lambda$)
lead to solutions that can be regarded as (stable) cosmological models.
}
 
\newpage
\section{Introduction} 

In a purely classical approach to ``strong
gravity", that is to say, in our geometric approach to hadron
structure (Recami \& Castorina 1976; Caldirola et al. 1978; Recami 1983;
Italiano 1984; Italiano \& Recami 1984), we came to associate hadron constituents with
suitable stationary, axisymmetric solutions of some new
Einstein--type equations, supposed to describe the strong field
inside hadrons. 
These Einstein--type equations are nothing but the
ordinary Einstein equations $ + \ \Lambda$ suitably
scaled down (Recami 1982; Recami \& Zanchin 1992). 
As a consequence, 
$ \Lambda$ and the masses $ M$ result, in such a theory, to be scaled up
and transformed into a ``hadronic constant" and into ``strong masses",
respectively.\\
Due to the unusual range of the values therefore assumed by
$\Lambda$, $ M$ and by other parameters (see $\S 2$), and even more
due to our requirements,
we met a series of solutions of the KNdS type, which had not received 
enough attention in the previous
literature. In particular, the requirement that those ``(strong) BH"
solutions be "{\it stable} ", i.e., that their surface temperature 
be vanishingly small (Recami et al. 1986; Recami \& Zanchin 1986; 
Zanchin 1987), implies the
coincidence of at least {\em two} of their (three, in general) horizons.
Aim of the present paper is putting forth for the first time such results,
while
``rephrasing" them in the more popular language of
ordinary gravity. \ Some of the more important points of originality of 
the present approach resides in our particular point of view; 
{\em i.e.}, in the fact that we are going to regard every black--hole
studied below as a (localized) object described by an {\em external}
observer living in a four-dimensional space-time asymptotically de 
Sitter (part of this work has been already deal by two of us, 
Zanchin et al. 1994).\\ 
The plane of the paper is as fellows: in $ \S 2$ we describe the 
main properties of the horizons associated with a central, stationary body 
and in $ \S 3$ we speak of the horizon temperature. In $ \S 4$ we deal 
both of the stable Schwarzschild--deSitter black hole and of the 
mass formulas for stable KNdS black holes while in $ \S 5$ we show the 
particular case of the triple coincidences. Finally in $ \S 6$ 
we report our conclusions.
 
\section{The horizons associated with a central, stationary body, and their
main properties}
  
We consider Einstein equations with cosmological term
\begin{equation}
R_{\mu \nu} - \frac{1}{2} g_{\mu \nu} R^{\rho}_{\rho} + \Lambda
g_{\mu \nu} = - kT_{\mu \nu} 
\end{equation}
where $ k \equiv \frac{8 \pi G}{c^{4}}$. We
choose, whenever convenient, units such that $G = 1$ and $c = 1$, and
look for the vacuum solutions describing the stationary axisymmetric
field created by a rotating charged source. This solution is the
KNdS space--time, whose metric in Boyer--Lindquist (1967)--
type coordinates $(t, r, \theta, \varphi)$ and being 
$m \equiv GM/c^{2}$ and $ a \equiv J/Mc,\;$ 
one can write:  
\begin{eqnarray}
ds^{2} & = & - \rho^{2} [dr^{2}/B + d \theta^{2}/D] - \rho^{-2} A^{-2}
    [(adt - (r^{2} + a^{2}) d \varphi]^{2} \sin^{2}\theta + \nonumber\\
& & \hspace*{0.3 cm} + BA^{-2} \rho^{-2} [dt - a \sin^{2}\theta \;
    d \varphi]^{2}
\label{eq:due}
\end{eqnarray}
where $\rho^{2}, A, \; B \equiv B(r), \; D \equiv D(\theta), \; Q^{2}$ are 
definited in Carter (1970, 1973) and the quantities $M, J$ and $q$ are
mass, angular momentum and electric charge of the source, respectively. For
simplicity, we here analyze only the case $\Lambda > 0$.\\
One meets the event horizons of the space Eq. (2) in
correspondence with the divergence of the coefficient $g_{rr}$, i.e., when
$B(r) = 0$. This equation,
\begin{equation}
(r^{2} + a^{2})(1 - \frac{\Lambda r^{2}}{3}) - 2mr + Q^{2} = 0 
\label{eq:tre}
\end{equation}
admits four roots, one of which, $r_{0}$, is always real and negative.
The interesting case is when Eq. (3) has four real solutions; in that
case we shall have {\em three} positive roots: let us call them
$r_{1}, r_{2}, r_{3}$, with $r_{3} \geq r_{2} \geq r_{1}$. 
We shall see that at $r = r_{3}$ we have a
cosmological horizon Carter (1970, 1973), while at $r = r_{2}$ and 
$r = r_{1}$ we meet two BH horizons analogous to the two wellknown 
$r = r_{+}$ and $r = r_{-}$ horizons of the Kerr metric.\\
The three horizons 1, 2, 3, in the general case when
they are all real, divide the space in the four parts I, II, III and
IV (see Fig. 1). On each horizon, quantity $g_{rr} \equiv g_{11}$ diverges,
i.e. $g^{rr} = 0$. 
In regions III and I it is always $g^{rr} < 0$, as expected in the case 
of an ordinary Kerr BH; on the contrary, in regions II and IV it is 
always $g^{rr} > 0$.\\

\begin{figure}[ht]
\psfig{file=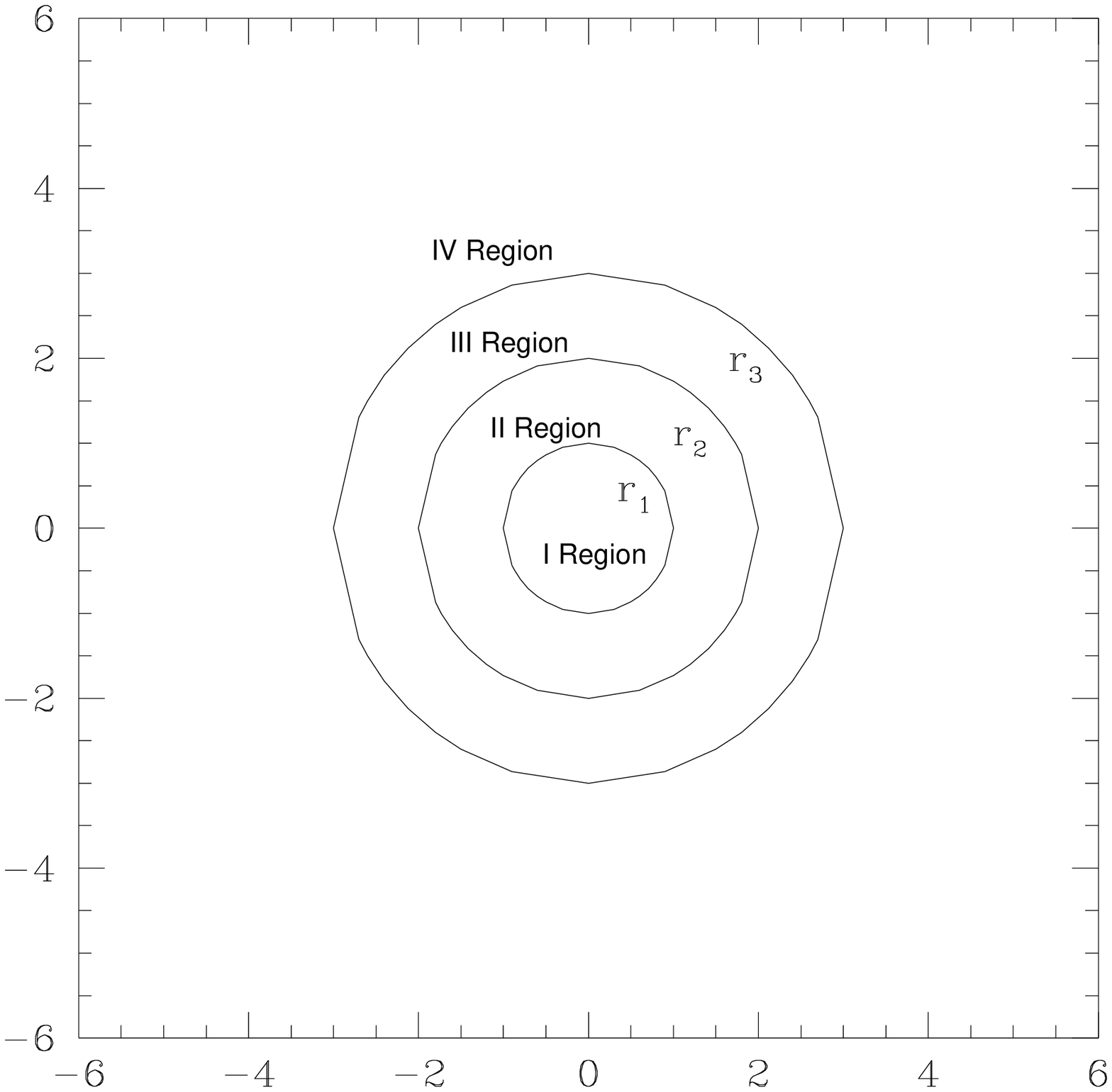,width=10cm}
\caption[h]{Given an, almost, pointlike stationary body, generating, when
$\Lambda \neq 0$, a KNdS space--time, {\em in the
Boyer--Lindquist coordinates} it will in
general possess {\em three} horizons 1,2,3, which divide the associated
space into the four regions I, II, II, IV. Surface 3 is the
cosmological horizon while surfaces 2,1 are the outer and inner BH
horizons, respectively.}
\end{figure}
Actually, it is possible to define a Killing vector 
$K^{\mu}$ which is simultaneously time--like in regions III and I but not 
in regions II and IV too (Recami 1978; Recami \& Shah 1979; 
Pav\u{s}i\u{c} \& Recami 1982; Recami \& Rodrigues Jr. 1982; 
Italiano 1986; Recami 1986; Trofimenko \& Gurin 1987). Therefore one 
can have stationary observers ($r =$ const.) only in 
regions III and I, in the sense that only there the  ($r =$
const.)  trajectories are time--like. We call {\em time--like}
the (ordinary type) {\em regions} III and I; and {\em space--like} the
other two {\em regions}.
For simplicity's sake, let us confine ourselves to the
static case (Reissner--Nordstr\"{o}m--de Sitter metric), which is not
qualitatively different. From Eq. (2), by putting $a = 0$, we find for
those geodesics:
\begin{equation}
ds^{2} = F dt^{2} - F^{-1} dr^{2} = 0 
\label{eq:qua}
\end{equation}
where $ F \equiv B/r^{2}$. By integration of Eq. (4), after some algebra 
one gets:
\begin{equation}
t = \mp 3 \Lambda^{-1} \displaystyle{\sum^{3}_{m=0}} \alpha_{m}
r^{2}_{m} \log | \displaystyle{\frac{r}{r_{m}}} - 1 | + C_{\mp}\;\;\;, \;\;\;
[m = 0,1,2,3]
\label{eq:cin}
\end{equation}
where $C_{\mp}$ are integration constants, and $\alpha_{m}$ are
``constants" whose value depends on the values of the four roots
$r_{0}, r_{1}, r_{2}, r_{3}$ of Eq. (3). In Eq. (\ref{eq:cin}) the 
upper (lower) sign corresponds to outgoing (ingoing) geodesics.\\
\h The behaviour of the radial null
geodesics $t = t(r)$ is given in Fig. 2 for the four regions.  One has
to recall that, however, such a figure does {\em not} represent in a
complete manner
the causal structure of our space-time.  That structure can be inferred
from the Penrose--Carter diagram (Carter 1966; Hawking \& Ellis 1973; 
Gibbons \& Hawking 1977; Mellor \& Moss 1989; Davies 1989).
\begin{figure}[ht]
\psfig{file=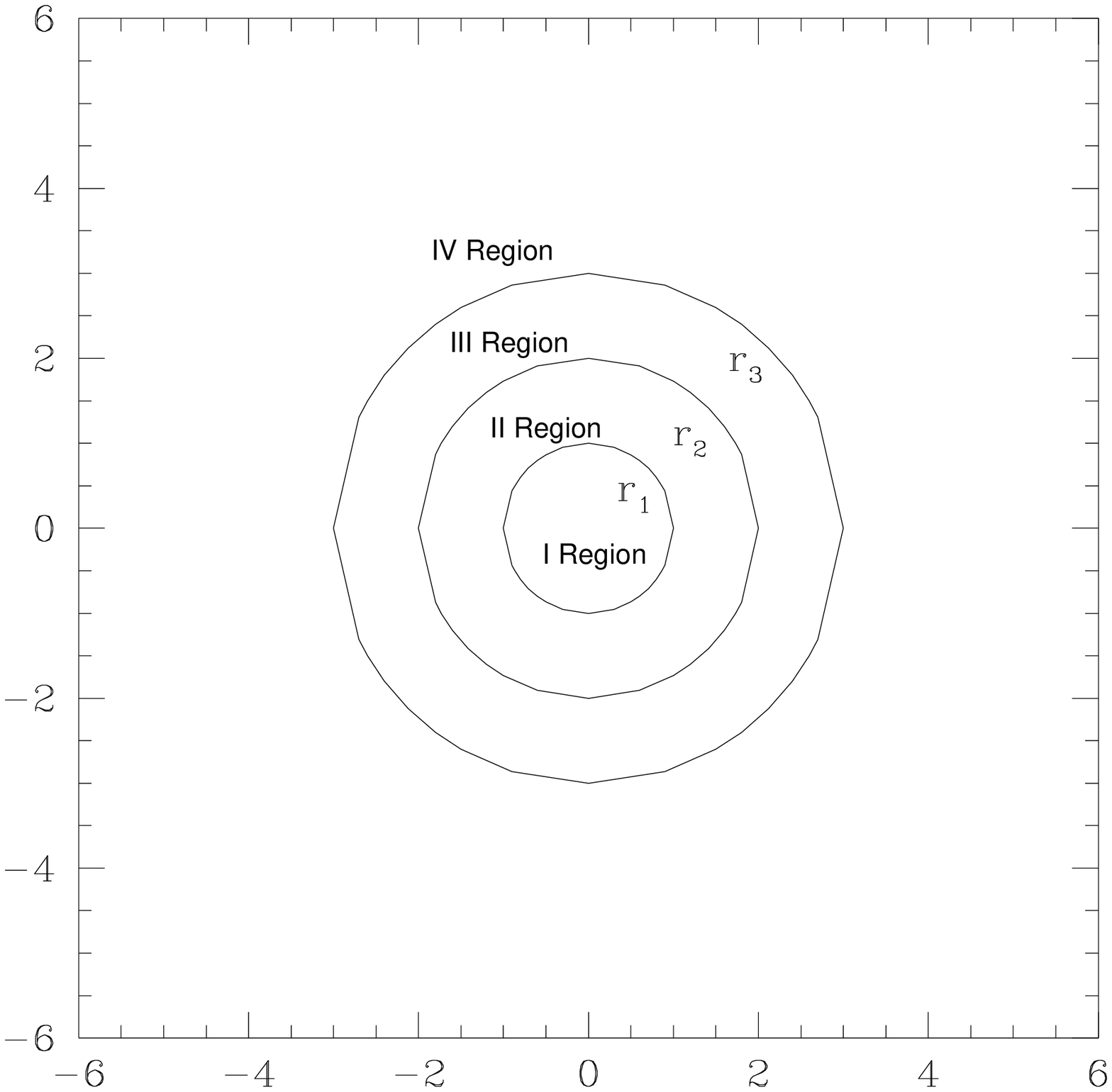,width=10cm}
\caption[h]{Behaviour of the radial null
geodesics in the different four regions of the Reissner--Nordstr\"{o}m--de
Sitter geometry (static case), {\em in Schwarzschild--like coordinates}.
In this case we have $r_{2} = 2r_{1} \;\; ; \;\; r_{3} = 3r_{1}$
(with $r_{0} = - r_{1} - r_{2} - r_{3}$).}
\end{figure}
 
 
\section{On the horizon temperatures}
 
In the general (stationary, i.e. KNdS)
case of metric Eq. (2), the Bekenstein--Hawking temperature $T_{n}$ 
(Bekenstein 1974; Hawking 1975) 
of each horizon in Fig. 1, is known to be proportional to the
horizon surface gravity as follows:
\begin{equation}
T_{n} = \varepsilon \gamma_{n}  \; , \;\;\;\;\;\; [n = 1,2,3]
\label{eq:sei}
\end{equation}
where $ \varepsilon \equiv \hbar/(2 \pi k_{B} c)$. On any
null--surface and in particular on every horizon the surface
gravity (Bardeen et al. 1973; Zheng \& Yuanxing 1983) can be defined 
by the equation
\begin{equation}
\partial_{\mu} (K_{\nu} K^{\nu}) = - 2 \gamma K_{\mu}
\label{eq:set}
\end{equation}
where the symbols $\partial_{\mu}$ representing the covariant derivatives. 
To evaluate the surface gravities $\gamma_{n}$, 
we here want remind everbody that in our metric (see Eq. (2)) does admit 
two Killing vector $ K^{\mu}_{t}, K^{\mu}_{\varphi}$, such that one 
can construct Killing vectors $ K^{\mu}_{n} = K^{\mu}_{t} + 
\omega_{n} K^{\mu}_{\varphi}$. Then we
finally get for the horizon
temperatures (and for $\Lambda \neq 0$) the expressions $T_{n} =
\varepsilon \gamma_{n}$ with
\begin{equation}
T_{n} = \frac{\varepsilon \Lambda}{6A(r^{2}_{n} + a^{2})}
\cdot \biggl| \Pi_{\ell \neq n}^{0,3} (r_{n} - r_{\ell}) \biggr| \;\;, \;\;\;\;
[\ell = 0,1,2,3]
\label{eq:ot}
\end{equation}
Eq. (\ref{eq:ot}) yields the result that the horizon temperature
can be vanishing small only when two or more horizons tend to coincide;
i.e., when two (or more) roots $r_{i}$ of Eq. (3) tend to coincide.
This result 
implies some {\em relations} among mass, radius, charge, angular 
momentum and $\Lambda$ of a stable BH.\\
In the particular (Kerr--Newman) case when $\Lambda = 0$, one gets only two 
(or no) horizons, corresponding to $r_{\pm} = m \pm \sqrt{m^{2} - a^{2} 
- Q^{2}}$, and Eq. (\ref{eq:ot}) has to be replaced by 
$T_{\pm} = \varepsilon(r_{+} - r_{-}) / (r_{\pm}^{2} + a^{2}$). 
We get a stable ($T = 0$) BH solution when
\begin{equation} 
r_{+} = r_{-} = m 
\label{eq:nov}
\end{equation}
 
that is to say, when the RL condition does hold:
  
\begin{equation} 
m^{2} = a^{2} + Q^{2}
\label{eq:die}
\end{equation}
 
However, since in this case region II disappeared, then the whole
BH--interior is time--like!, so as the external region III.
We call a solution of this type a
{\em ``time--like black--hole"}. 
 

\section{Particular cases of stable BH}

An interesting case is that of the Schwarzschild--de
Sitter metric (Gibbons \& Hawking 1977), in which
$Q^{2}=a^{2}=0$, so that $B = - \Lambda r^{4}/3 + r^{2} - 2mr$ and 
two horizons only (with radii $r_{-} \equiv
r_{\scriptstyle{{\rm B}}} \;,\; r_{+} \equiv r_{\scriptstyle{{\rm C}}}$,
respectively) are met, whose surface temperatures result to be
\begin{equation}
T_{\pm} = \frac{\varepsilon \Lambda}{3 r_{\pm}^{2}} (\frac{3m}{\Lambda}
- r^{2}_{\pm})
\label{eq:und}
\end{equation}
Once more, the requirement $T = 0$ implies that $r_{\scriptstyle{{\rm B}}} =
r_{\scriptstyle{{\rm C}}} \equiv r$
and that $r = (3m/ \Lambda)^{1/3}$. The last equation can be read as
\begin{equation}
r = \Lambda^{- 1/2} = 3m \;\;\; , \;\;\;
[r_{\scriptstyle{{\rm B}}} = r_{\scriptstyle{{\rm C}}} \equiv r]
\label{eq:dod}
\end{equation}
since those two radii coincide, only, when
\begin{equation}
9 \Lambda m^{2} = 1
\label{eq:tred}
\end{equation}
More interesting, here, is the observation that $r_{-}$
and $r_{+}$ behave so as $r_{2}$ and $r_{3}$, respectively, of Figs. 1 and 2.
For this reason we called $r_{-} \equiv r_{\scriptstyle{{\rm B}}}$ and 
$r_{+} = r_{\scriptstyle{{\rm C}}} \;$  where B $\equiv$ {\it BH horizon} 
 and C $\equiv$ {\it cosmological horizon}. 
When $r_{\scriptstyle{{\rm B}}}$ tends to coincide with
$r_{\scriptstyle{{\rm C}}}$,
the {\it time--like} regions of type III do disappear, so that we are left only with
regions of type II and IV, and the BH tends to occupy the whole space
inside the cosmological horizon (roughly speaking, the BH itself can
be regarded as a model for a cosmos). It is worthwhile mentioning that,
by choosing for
$\Lambda$ the value $| \Lambda| \approx 10^{-52}$ m$^{-2}$ ordinarily
assumed for our cosmos, the condition given by Eq. (\ref{eq:tred}) 
yields $M \simeq \frac{1}{2} \times 10^{53}$ kg, which is close 
to the estimated mass of our own cosmos.\\
Now, we consider the characteristics of stable BHs
in the general (KNdS) case when the source is
endowed also with $ J$ (stationary case) and $ q$. We have
at our disposal two equations: 
\begin{equation}
\left\{ \begin{array}{l}
B(r) = - \displaystyle{\frac{\Lambda r^{4}}{3}} + (1 - \displaystyle{\frac{\Lambda
a^{2}}{3}}) r^{2} - 2mr + a^{2} + Q^{2} = 0\\
\\
T = - 2 \displaystyle{\frac{\Lambda r^{3}}{3}} + (1 - \displaystyle{\frac{\Lambda
a^{2}}{3}}) r - m = 0
\end{array} \right.
\label{eq:quatto}
\end{equation}
the second equation requires the vanishing of the derivative $B'(r)$ in
correspondence with the values $r_{n}$ which satisfy the first equation
$[B(r) = 0]$. Such second
equation, therefore, ensures the solutions of the system to be double
(or triple) ``roots" of eq. $B(r) = 0$.\\
After, some algebra, we get explicitly 
a second equation providing
us with a link among the various parameters $m, \Lambda, a, Q$:
\begin{equation}
\left\{ \begin{array}{l}
r = \displaystyle{\frac{3 m \sigma}{E}}\\
\\
9 m^{2} \sigma(\delta \sigma - E) + 2 \eta E^{2} = 0
\end{array} \right.
\label{eq:quin}
\end{equation}
where $ E = 3 \delta^{2} + 4 \Lambda \delta \eta - 18 m^{2} \Lambda$
being
\begin{eqnarray}
\delta  & = & 1 - \Lambda a^{2}/3  \nonumber \\
\eta  & = & a^{2} + Q^{2} \nonumber \\
\sigma & = & \delta^{2} - 4 \Lambda \eta \nonumber
\end{eqnarray}
Eqs. (\ref{eq:quin}) do yield, of course, {\em both} the stable BH solutions
resulting from the coincidence of $r_{1}, r_{2}$, {\em and} those
resulting from the coincidence of $r_{2}, r_{3}$. Namely, the second
of Eqs. (\ref{eq:quin}) can be written as
\begin{equation}
\frac{2 \delta \sigma}{E} = 1 \pm
\sqrt{1 - \frac{8 \delta \eta}{9 m^{2}}} 
\label{eq:sedi}
\end{equation}
from which one can, of course, construct two independent systems.\\
We, starting, consider the case when $r_{1} \equiv r_{2} 
= r_{-}$. In this case the regions of type II (Figs. $ 1 \div 2 $) do
disappear and we obtain a {\em stable} Kerr--Newman--de
Sitter BH, similar to the {\em stable} BH encountered in $ \S 3$, 
in the particular Kerr--Newman case. In that case,
however, the stable BH was surrounded by asymptotically flat regions
of type III; whilst in the present case (we are still in presence of 
a cosmological $r = r_{3}$ horizon) our stable BH is surrounded by
{\em two} type regions: regions of type III, and regions of type IV. 
In other words, both the external (III) and the
internal (I) regions of the present stable BH are time--like regions,
separated just by a semi--permeable membrane.
In these regions, any causal observer $O_{\rm c}$ can live therein 
without falling into the singularity $r = 0$.\\
Now, we consider the case when $r_{2} \equiv r_{3} = r_{+}$. In this second 
case the regions of type III (see Figs. $ 1 \div 2 $) do
disappear and we obtain a BH, bounded by a {\em stable}
horizon originating from the fusion of a BH--type
($r_{2}$) surface and a cosmological--type ($r_{3}$) horizon. The
stable $r_{2} \equiv r_{3}$ null--surface can be regarded, therefore,
both as a BH--membrane and as a cosmological horizon. Outside such a
surface, we meet regions of type IV, asymptotically de Sitter. 
Both the internal (II) and the external (IV) BH
regions are space--like, since the time--like type III regions, 
where causal observers usually live, disappeared. In regions II and IV
no stationary observers can exist. 
Inside the $r_{2} \equiv r_{3}$ surface, we moreover
find at $r = r_{1}$ a null surface that can be considered the internal
BH boundary, so as in the Kerr--Newman (or Kerr) case. 
 
 
\section{Triple coincidences}

In the very special case when all the three positive
roots of Eq. (3) do coincide, i.e. when $r_{1} = r_{2} = r_{3}$, we
shall meet a stable BH with a single horizon, whose radius takes on a
simple analytical expression. Let us write Eq. (\ref{eq:sedi}), 
after some algebra, more conveniently, as:
\[
r = \frac{3 m}{2 \delta} \pm \sqrt{\frac{9 m^{2}}{4 \delta^{2}} -
\frac{2 \eta}{\delta}}
\]
and observe that the condition of triple coincidence 
requires the vanishing of the square root, i.e. yields the solution:
\begin{equation}
r = \frac{3 m}{2 \delta} 
\label{eq:dicias}
\end{equation}
with the two simultaneous RL constraints: 


\begin{equation}
m^{2} = \displaystyle{\frac{8}{9}} \delta(a^{2} + Q^{2}) 
\label{eq:diciot}
\end{equation}

\begin{equation}
m^{2} = \displaystyle{\frac{2}{9}} \displaystyle{\frac{\delta^{3}}{\Lambda}}
\label{eq:dician}
\end{equation}

Eq. (\ref{eq:dician}) comes from inserting Eqs. (\ref{eq:dicias}), 
(\ref{eq:diciot}) in either of Eqs. (\ref{eq:quatto}).\\
In the present case, all the regions II and III did disappear;
and the " type I " interior of our stable BH is time--like whilst its 
" type IV " exterior is space--like.  Such solution is therefore a
{\em ``time--like black--hole"}. Again, regions IV are asymptotically de
Sitter. 
Such a BH solution is conveniently interpretable (see $\S 4$) 
also as a cosmological model: namely, as a model of a {\it stable} cosmos.
 

\section{Conclusion} 

Let us stress, first of all, that for stable BHs we got 
`` RL " relations among their $ m$, $ J$, $ q$ and 
$ \Lambda$.
For instance, in the case $\Lambda = 0$ we got Eq. (\ref{eq:die}) which, 
when $ q$ is negligible, can just be written if we assume like units 
$ G = 1$ and $ c = 1$ as
 
\vspace*{0.5 cm}
 
\hfill{$M^{2} = J$.
\hfill} ($10^{\star}$)
 
\vspace*{0.5 cm}
 
On the contrary, when $J = 0$ and $ q$ is still negligible, then we meet Eq.
(\ref{eq:tred}), which, always with $ G = 1$ and $ c = 1$, can read

\vspace*{0.5 cm}
 
\hfill{$M^{2} = \displaystyle{\frac{1}{9}} \Lambda^{-1}$.
\hfill} ($13^{\star}$)
 
\vspace*{0.5 cm}

In the most general case, the considered relation (among $M, J, q,
\Lambda$) is involute, and was given by the second one of Eqs. (\ref{eq:quin}). In the simpler case of $ \S 6$, i.e. of the ``triple coincidence", we
obtained two such relations, namely Eqs. (\ref{eq:diciot}), 
(\ref{eq:dician}), which are still complicated. However, if 
$| \Lambda a^{2}| \ll 1$, Eqs. (\ref{eq:diciot}), (\ref{eq:dician}) 
yield both 
 
\vspace*{0.5 cm}
 
\hfill{$m^{2} \simeq \displaystyle{\frac{8}{9}} (a^{2} + Q^{2}) \;$ ,
\hfill} (20)
 
\vspace*{0.5 cm}
 
to be compared with Eq. (\ref{eq:die})  and 
 
\vspace*{0.5 cm}
 
\hfill{$M^{2} \simeq \displaystyle{\frac{2}{9}} \Lambda^{-1}\;$ ,
\hfill} (21)
 
\vspace*{0.5 cm}
 
that if $ c = G = 1$ can be compared with Eq. ($13^{\star}$).\\
The most interesting point is that, with the exception of
Eqs. (\ref{eq:die}), ($10^{\star}$), all such ``RL" relations 
can be attributed also to our stable cosmological models, i.e., to our 
stable {\em ``cosmoses"}. 
Finally, let us mention that elsewhere we shall apply
and interpret the results presented in this paper to the case of ``strong
gravity" theories and ``strong BHs": i.e., to the case of hadronic
physics.
 
\vspace*{1 cm}
 
{\bf Acknowledgements:} Useful discussions are acknowledged with M. Bordin,
A. Bugini, R.H.A. Farias, R. Garattini, E. Giannetto, Italiano A..,
G.D. Maccarrone, E. Majorana jr., E.C. de Oliveira, N. Paver,
G. Salesi, C. Sivaram, Q.A.G. de Souza, and particularly with P. Ammiraju,
P. Bandyopadhyay, R.B. Campbell, V. De Sabbata, A. Insolia,
A. van der Merwe, S. Mukherjee, H.C. Myung, R. Pucci, W.A. Rodrigues Jr.,
M. Sambataro and S. Sambataro; as well as the kind collaboration of F.
Aversa, F. Nobili, M.T. Vasconselos and M. Zorzini.
Special thanks go to Yuval Ne'eman, J.A.Roversi and L.B.Annes. 
 


\begin{thebibliography}{11}
\bibitem{} Bardeen J.M., Carter B. and Hawking S.W., Comm. Math.
Phys. {\em 31} (1973) 1161
\bibitem{} Bekenstein J.D., Phys. Rev. {\em D9} (1974) 3292
\bibitem{} Boyer R.H. and Lindquist R.W., Jour. Math. Phys. {\em 8}
(1967) 265
\bibitem{} Caldirola P., Pav\u{s}i\u{c} M., Recami E., Nuovo Cimento {\em B48} (1978) 205
\bibitem{} Carter B., Phys.Rev. {\em 141} (1966) 1242,
in particular Fig.1 therein
\bibitem{} Carter B., Comm. Math. Phys. {\em 17} (1970) 233
\bibitem{} Carter B., in {\em ``Les Astres Occlus"} (Gordon and Breach; N.Y., 1973)
\bibitem{} Davies P.C.W., Class.Quantum Grav.
{\em 6} (1989) 1909
\bibitem{} Gibbons G.W. and Hawking  S.W., Phys.Rev. {\em D15} (1977) 2738,
in particular Fig.4 therein
\bibitem{} Hawking  S.W., Comm. Math. Phys. {\em 43} (1975) 199
\bibitem{} Hawking  S.W. and Ellis G.F.R., {\em ``The
Large Scale Structure of Space--Time"} (Cambridge Univ. Press, 1973),
in particular Sects. 5.5, 5.6; and references therein
\bibitem{} Italiano A., Hadronic J. {\em 9} (1986) 9
\bibitem{} Italiano A., {\em et al.}, Hadronic J.,  {\em 7} (1984) 1321
\bibitem{} Italiano A. and Recami E., Lett. Nuovo Cimento {\em 40} (1984) 140
\bibitem{} Mellor F. and Moss I., Class.Quantum Grav. {\em 6} (1989) 1379,
in particular Fig.1 therein
\bibitem{} Pav\u{s}i\u{c} M. and Recami E., Lett. Nuovo Cim. {\em 34}
(1982) 357
\bibitem{} Recami E., Found Phys. {\em 8} (1978) 329
\bibitem{} Recami E., Prog. Part. Nucl. Phys. {\em 8} (1982) 401
\bibitem{} Recami E., Found. Phys. {\em 13} (1983) 341
\bibitem{} Recami E., Riv. Nuovo Cimento {\em 9} (1986), issue
no. 6
\bibitem{} Recami E. and Castorina P., Lett. Nuovo Cimento {\em 15} (1976) 357
\bibitem{} Recami E. and Rodrigues Jr. W.A., Found. Phys. {\em 12}
(1982) 709
\bibitem{} Recami E. and Shah K.T., Lett. Nuovo Cimento
{\em 24} (1979) 115
\bibitem{} Recami E. and Zanchin V.T., Phys. Lett. {\em B177} (1986) 304
\bibitem{} Recami E. and Zanchin V.T., Il Nuovo Saggiatore {\bf 8} (1992, issue
no.2) 13
\bibitem{} Recami E., Mart\'{\i}nez J.M. and Zanchin V.T., Prog. 
Part. Nucl. Phys. {\em 17} (1986) 143
\bibitem{} E.Recami, V.T.Zanchin, J.A. Roversi \& L.B. Annes: ``Regge--like
Relations for non-evaporating Black-holes",\hfill\break
{\em Found. Phys. Letters} {\bf 7} (1994) 167-179.\hfill\break
\bibitem{} Trofimenko A.P. and Gurin V.S., Pram\~{a}na {\em 28}
(1987) 379
\bibitem{} Zanchin V.T., M. Sc. thesis (UNICAMP, 1987)
\bibitem{} Zanchin V.T., Recami E., J.A. Roversi, L.A. Brasca-Annes, Found. 
of Phys. Lett. {\em 7} (1994, issue no. 2) 167
\bibitem{} Zheng Z. and Yuanxing G., {\em Proc.
3$^{\underline{rd}}$ Marcel Grossmann Meeting on Gen. Relat.,} ed. by
H. Ning (North--Holland; Amsterdam, 1983),  p. 1177
\end{thebibliography}
\end{document}